\documentclass[12pt,preprint]{aastex}

\def\deg{\ifmmode^\circ\else$^\circ$\fi}

\def\mic{~$\mu$m}

\def\mic{$\mu${\rm m}}
\def\lir{$L_{\mbox{IR}}$}

\def\arcs{\ifmmode {''}\else $''$\fi}
\def\arcm{\ifmmode {'}\else $'$\fi}
\def\parcs{\sa=.07em \sb=.03em
     \ifmmode $\rlap{.}$^{\scriptscriptstyle\prime\kern -\sb\prime}$\kern -\sa$
     \else \rlap{.}$^{\scriptscriptstyle\prime\kern -\sb\prime}$\kern -\sa\fi}
\def\parcm{\sa=.08em \sb=.03em
     \ifmmode $\rlap{.}\kern\sa$^{\scriptscriptstyle\prime}$\kern-\sb$
     \else \rlap{.}\kern\sa$^{\scriptscriptstyle\prime}$\kern-\sb\fi}

\def\spose#1{\hbox to 0pt{#1\hss}}
\def\simlt{\mathrel{\spose{\lower 3pt\hbox{$\mathchar"218$}}
     \raise 2.0pt\hbox{$\mathchar"13C$}}}
\def\simgt{\mathrel{\spose{\lower 3pt\hbox{$\mathchar"218$}}
     \raise 2.0pt\hbox{$\mathchar"13E$}}}
\def\lsim{\rlap{$<$}{\lower 1.0ex\hbox{$\sim$}}}
\def\gsim{\rlap{$>$}{\lower 1.0ex\hbox{$\sim$}}}

\begin{document}

\title{16 micron Imaging around the Hubble Deep Field North with the {\it Spitzer}$^1$\ IRS$^2$
}

\altaffiltext{1}{based on observations obtained with the {\it Spitzer Space Telescope}, which is 
operated by JPL, California Institute of Technology for the National Aeronautics and Space Administration}

\altaffiltext{2}{The IRS is a collaborative venture between Cornell
 University and Ball Aerospace Corporation that was funded by NASA
   through JPL.}

\author{H. I. Teplitz\altaffilmark{3}, 
V. Charmandaris\altaffilmark{4,5,6},
R. Chary\altaffilmark{3},
J. W. Colbert\altaffilmark{3},
L. Armus\altaffilmark{3},
D. Weedman\altaffilmark{4}
}

\altaffiltext{3}{Spitzer Science Center, MS 220-6, Caltech, Pasadena, CA 91125.  hit@ipac.caltech.edu}
\altaffiltext{4}{Astronomy Department, Cornell University, Ithaca, NY  14853}
\altaffiltext{5}{Chercheur Associ\'e, Observatoire de Paris, F-75014, Paris, France}
\altaffiltext{6}{Univ. of Crete, Dept. of Physics, GR-71003 Heraklion, Greece}


\begin{abstract}
  
  We present a pilot study of 16 $\mu$m imaging within the GOODS
  northern field.  Observations were obtained using the PeakUp imaging
  capability of the {\it Spitzer}\ IRS.  We survey 35 square
  arcminutes to an average 3$\sigma$\ depth of 0.075 mJy\ and detect
  149 sources.  The survey partially overlaps the area imaged at 15
  $\mu$m by ISO, and we demonstrate that our photometry and
  galaxy-number counts are consistent with their measurements.  We
  infer the total infrared luminosity of 16 \mic\ detections using a
  comparison to local templates and find a wide range of $L_{\rm IR}$\ 
  from $\sim 10^9$\ to $10^{12}$\ $L_{\odot}$.  Approximately one fifth of the
  detected sources have counterparts in the {\it Chandra}\ 2 Msec
  catalog, and we show that the hard band (2-8 keV) detected sources
  are likely to have strong AGN contributions to the X-ray flux.  The
  ultradeep sensitivity of {\it Chandra}\ implies some X-ray
  detections may be purely starbursting objects.  We examine the 16 to
  24 $\mu$m flux ratio and conclude that it shows evidence for the
  detection of redshifted PAH emission at $z\sim 0.5$ and $z>0.8$.

\end{abstract}


\section{Introduction}

The {\it Spitzer Space Telescope}\ offers unique sensitivity in the
mid-infrared (MIR) for the study of star formation in distant
galaxies.  Photometric measurement of the spectral energy distribution
(SED) of such sources is a powerful tool, but the wavelength gap
between 8 \mic\ in IRAC channel four\citep{Fazio 2004}\ and 24 \mic\ 
in MIPS \citep{Rieke 2004}\ limits the study of galaxies at $z\sim 1$,
where prominent features fall in between the wavelength coverage of
the two instruments.  The wavelength gap can be filled with
observations using the {\it Spitzer}\ IRS blue PeakUp filter, which
samples wavelengths from 13 to 18.5 \mic\ \citep{Houck 2004}.

Starting in the second year of operations, a science quality PeakUp
Imaging (PUI) mode has been made available.  PUI observations allow us to
observe the evolution of mid-infrared (MIR) spectral features such as
6.2 and 7.7 $\mu$m PAH emission and the 9.7 $\mu$m silicate absorption
trough.  We can follow the 7.7 $\mu$m PAH feature from the local
universe where it lies in the IRAC channel 4 passband, to $0.8<z<1.3$\ 
where it falls in the 16 $\mu$m PUI filter, to redshifts 
near $z\sim 2$\ where it enters the 24 $\mu$m MIPS filter.
Similarly, the silicate absorption can dominate the 16 and 24 \mic\ 
filters at $z\sim 0.6$\ and $\sim 1.5$, respectively.  The ratio of
flux measured in these filters can thus detect the presence of MIR
features \citep{Takagi and Pearson 2005,Elbaz 2005}.

Furthermore, PeakUp Imaging mode observations can detect galaxies significantly
below the brightness limit of IRS spectroscopy ($\sim 0.5-1$\ mJy).
\cite{Houck 2004}\ estimate a 3$\sigma$\ sensitivity of $\sim$ 0.1 mJy
in two minutes for PUI in a low background region.  Luminous infrared
galaxies (LIRGs and ULIRGs; with luminosities greater than $10^{11}$\ 
and $10^{12}~L_{\odot}$\ respectively) at moderate and high redshift
are expected to be easily detected at this flux level.

We present a pilot study of the use of the 16 \mic\ PUI mode.  We
choose as our target the Great Observatories Origins Deep Survey
\citep[GOODS;][]{Dickinson 2004,Giavalisco 2004}\ northern field.
GOODS provides an unprecedented opportunity to study distant galaxies
across the available wavelength spectrum.  The northern field in
particular has been the subject of the deepest X-ray observation ever
taken, the 2 Msec {\it Chandra}\ deep field \citep{Alexander 2003},
and has been observed extensively by the {\it Hubble Space Telescope}\ 
including the original Hubble Deep Field North
\citep[HDF-N;][]{Williams 1996}.  GOODS also includes the deepest {\it
  Spitzer}\ observations at 3-8 \mic\ with IRAC and 24 \mic\ MIPS.

The current survey consists of a small region (35 square arcminutes)
within the GOODS northern field using 16 $\mu$m PUI imaging.  This
survey partially overlaps the existing {\it Infrared Space
  Observatory}\ (ISO) ultradeep survey at 15 $\mu$m \citep{Aussel
  1999}.  We will show that photometry in the new survey is consistent
with that obtained by ISO.  We present the observing strategy and data
reduction in Section 2.  We provide some discussion of the PUI mode in
the first year of {\it Spitzer}\ operations.  In Section 3, we present
the results and the catalog.  We discuss the implications of the
survey in Section 4, and provide a summary in Section 5.  Throughout,
we assume a $\Lambda$-dominated flat universe, with $H_0=70$\ km
s$^{-1}$\ Mpc$^{-1}$, $\Omega_{\Lambda}=0.7, \Omega_{m}=0.3$.

\section{Observations and Data Reduction}

Data were obtained as part of IRS calibration activities during
Science Verification (SV) and in parallel with a Guaranteed Time
Observation (GTO) program to obtain photometry of
submillimeter-selected sources in the field \citep{Charmandaris 2004}.
The SV observations were motivated to provide a direct comparison of
{\it Spitzer}\ IRS 16 $\mu$m imaging of individual targets previously observed
by ISOCAM \citep{Aussel 1999}.  The GTO program targeted seven objects
mostly outside the ISOCAM area, but within the GOODS field.  In each
case the full PUI field of view was imaged for each object at both 16
and 22 $\mu$m.  During the 22 $\mu$m imaging, 16 $\mu$m data was
obtained in parallel.  In total, we obtained 72 exposures, of 60
seconds each, for a survey integration time of 1.2 hours.

While the astronomical observing template (AOT) for ``sample up the
ramp'' (SUR) PeakUp imaging was not commissioned until the second {\it
  Spitzer}\ proposal cycle (GO2), a work-around was developed using
the standard spectroscopic staring-mode AOT.  By selecting the
short-low (SL) slit and properly offsetting the telescope, PeakUp
images were obtained of selected areas around the HDF-N.  This
technique was used for both the SV and GTO observations, and it
resulted in the sparse 16 $\mu$m map shown in Figure \ref{fig:
  coverage}.

IRS 16$\mu$m images were reduced with the standard pipeline at the
Spitzer Science Center (SSC; see chapter 7 of the Spitzer Observer's
Manual\footnote{http://ssc.spitzer.caltech.edu/documents/som/}).
Although pipeline calibrated data have had nominal low-background sky
images subtracted, some residual sky signal may remain.  We create
median sky images from near-in-time subsets of the data, after object
masking.  We register the sky subtracted images using the MOPEX
software provided by
SSC\footnote{http://ssc.spitzer.caltech.edu/postbcd/}.  Given the
sparseness of objects in the fields, we rely on the reconstructed
pointing to provide the registration.  The pointing (without the
refinement afforded by known stars in the IRAC and MIPS) is typically
good to $\sim 1^{\prime \prime}$.  The Point Spread Function (PSF) at
16$\mu$m\ has a full width at half maximum (FWHM) of $\sim 3.6$\ 
arcseconds, compared to the IRAC 8 \mic\ PSF of 1.98\arcs\ and the MIPS
24 \mic\ width of 5.4\arcs.

We identify sources and measure photometry with the SSC's APEX$^{8}$\ 
software, which includes point source fitting to measure the total
flux of the source, as well as source deblending.  We cross correlate
our list of detected sources with the much deeper IRAC Channel 3 (5.8
\mic) image of the field from GOODS (Dickinson et al.\ 2005; in prep),
allowing a maximum of 2\arcs\ separation.  Using the IRAC data ensures
that few, if any, spurious objects are included, allowing us to set
our detection threshold at a faint level.  The faintest IRAC
counterpart has a flux of 10 $\mu$Jy.  At that flux level, the
integrated source counts are $\sim 18$\ per square arcminute
\citep{Lacy 2005}, so the chance of random contamination within our
search radius should be less than 6\%.

The GOODS 24 $\mu$m data in the HDF-N reach an unprecedented point
source sensitivity limit of $<$20 $\mu$Jy (Chary et al. 2005; in prep).
More than 3000 sources are detected in 24$\mu$m over the entire field.
The IRS 16$\mu$m sources were matched to the 24$\mu$m catalog using a
3$\arcsec$ positional matching threshold. However, more than 90\% of
the sources (143/149) are matched to within 1$\arcsec$ and the few
outliers were inspected by eye. These were found to be attributable to
source blending or low signal/noise in the 16$\mu$m image.

The 16 $\mu$m photometric zeropoint was derived from repeated
observations of standard stars.  The photometry is referenced to a 10K
black body spectrum which represents the stellar calibrator.  This
referencing is the same as the MIPS calibration, but is different than
IRAC and ISOCAM, both of which tie their photometric systems to
spectra with $\nu f_{\nu}=const$.  The color correction between the
systems, however, is small ($<3$\%). There is a systematic uncertainty
due to flux calibration on the order of 6\%.  The photometric zero
point has been verified with comparion of the PUI data to IRS spectra,
which are independently calibrated using different standard stars and
a different set of templates.  In addition, cross calibration between
22 \mic\ PUI data and MIPS 24 \mic\ photometry of faint galaxies shows
$<10$\%\ residuals for most high significance sources; this difference
is within the range attributable to the difference in wavelength
coverage.

We estimate a point source sensitivity of 75 $\mu$Jy, $3\sigma$, in
120 seconds.  This is somewhat better than numbers previously reported
\citep{Houck 2004}, and we attribute the difference to improvements in
the SSC pipeline, the very low background of the HDF region and the
large number of frames available to optimize the sky subtraction.

\section{Results}

We detect 149 objects, with fluxes ranging from 21 $\mu$Jy to 1.24
mJy.  From the literature, 90 of these sources have redshifts
\citep{Cohen 2000, Cowie 2004, Wirth 2004}, ranging from 0.11 to 2.59.
The median redshift is 0.7; only two sources lie at $z>1.5$.  All sources
have optical counterparts in the GOODS catalog.  Table 1
presents the photometry of the detected sources.

There are 100 {\it Chandra}\ sources from the 2 Msec catalog
\citep{Alexander 2003}\ within the surveyed area.  Of these, 33 have
16$\mu$m counterparts, comprising 22\% of the 16$\mu$m sample.
\cite{Fadda 2002}\ find a higher fraction (16 of 42) of {\it Chandra}\ 
counterparts to ISOCAM sources, but our survey extends to fainter MIR
fluxes.  The {\it Spitzer}\ survey area also covers 43 radio sources
from the 1.4 GHz survey of \cite{Richards 2000}, 24 of which have 16
\mic\ counterparts.  Of the these, 17 have {\it Chandra}\ detections
with 1\arcs\ of the radio position, and another 5 have {\it Chandra}\ 
within 2\arcs.  We defer more discussion of radio sources to future
papers, but see \cite{Marcillac 2005}\ for a comparison of these
sources with the ISOCAM survey.

We detect 24 objects in common with ISOCAM, including all of the high
significance sources from \cite{Aussel 1999}\ that fall in the {\it
  Spitzer}\ survey area.  Eight sources from the less significant ISO
sample are not detected by {\it Spitzer}.  Three ISOCAM sources are
resolved into multiple sources by the higher spatial resolution of the
PUI mode.  In addition, 11 {\it Spitzer}\ sources are not in the Aussel et
al.\ catalog, but fall within the central area of the ISOCAM map.
However, 6 of these have fluxes below 0.1 mJy.  Several additional
sources are undetected at the edges of the ISOCAM area, where the
coverage is less deep.  

The redshift distribution of the {\it Spitzer}\ and ISOCAM sources are
similar.  Matching the Aussel et al. ISOCAM source list to the
redshift catalog of \cite{Cowie 2004}\ yields 49 objects with spectroscopic
redshifts, with a median of $z=0.79$, compared to the {\it Spitzer}\ 
median of 0.7.  The shape of the redshift distributions is nearly
identical, with a steep roll off beyond $z\sim 0.8$\ and strong
detections in the known redshift overdensities at $z\sim 0.45$\ and
$z\sim 0.8$\ \citep{Cohen 2000}. 

\begin{figure}[t*]

\plotone{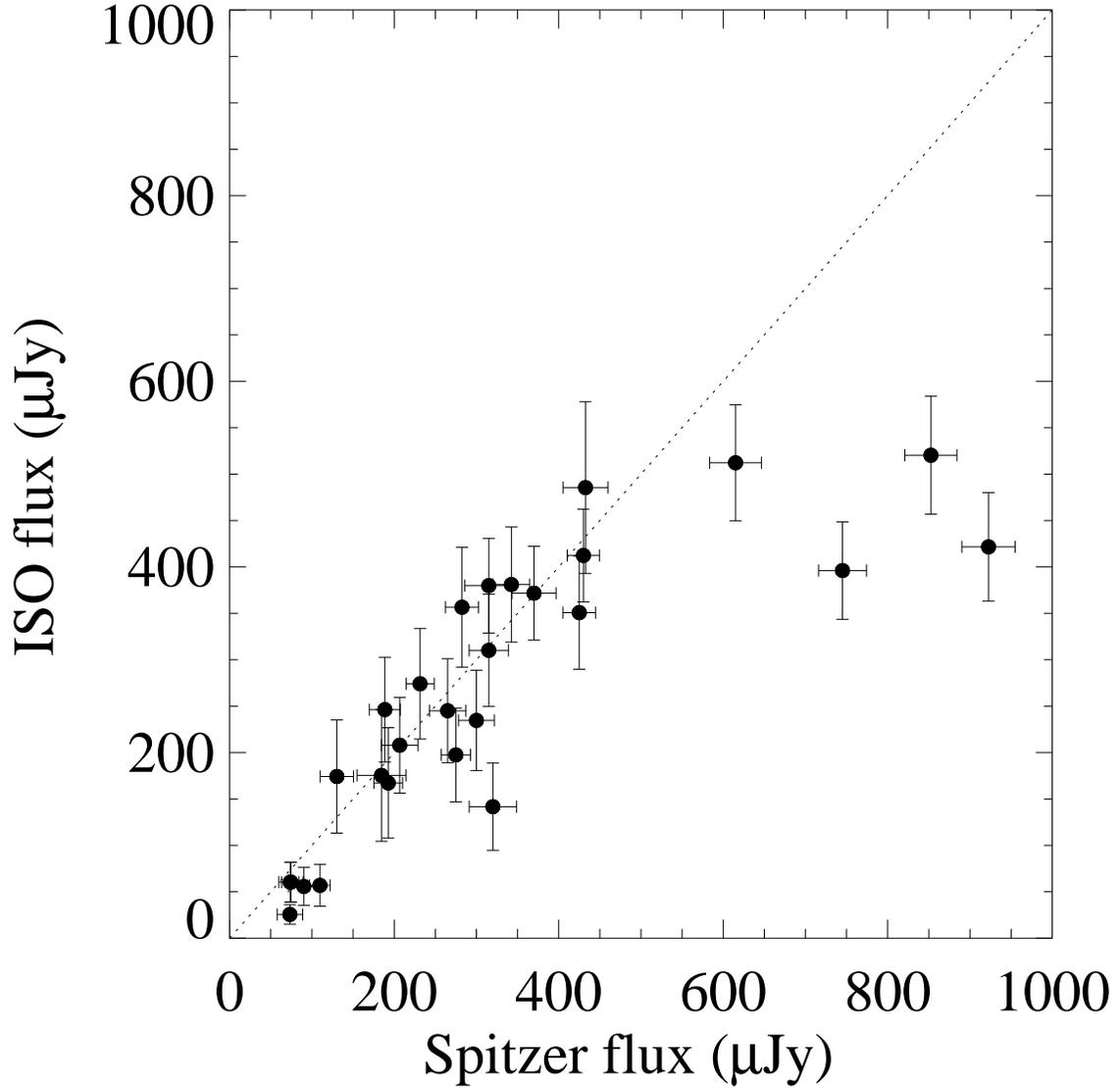}
\caption{\label{fig: iso compare}
The flux measured with the {\it Spitzer}\ IRS in the 16 $\mu$m bandpass is compared to that 
measured with ISOCAM in the
15$\mu$m bandpass (LW3) for the objects commonly 
detected.  The diagonal line indicates equal flux.  Error bars indicate
the 1$\sigma$\ uncertainties.  Fluxes have been color corrected to the 
{\it Spitzer}\ photometric system, and for differences between the IRS 16 $\mu$m 
and ISOCAM LW3 bandpasses (see text).
  }

\end{figure}

\begin{figure}[t*]

\plotone{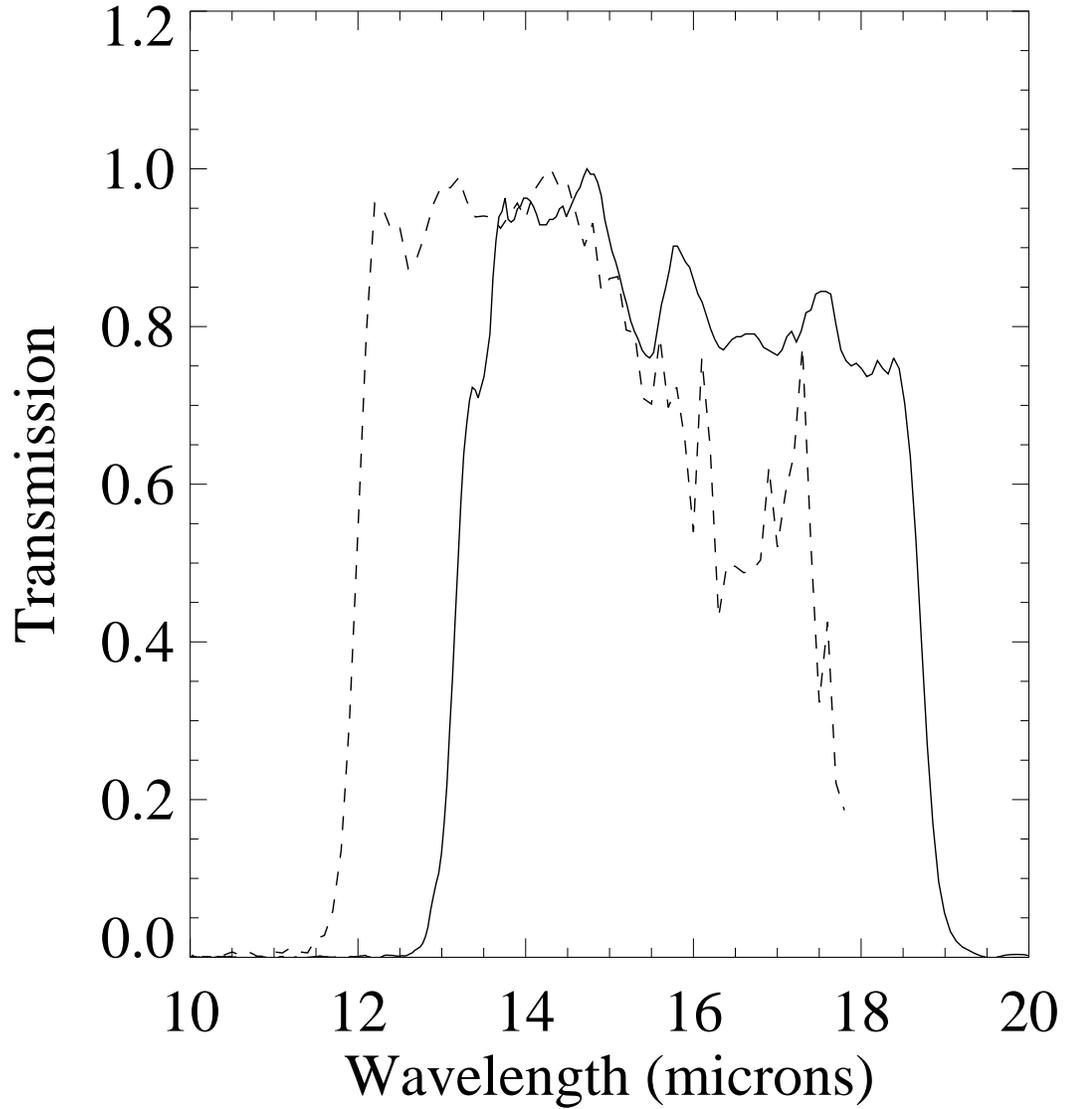}
\caption{\label{fig: pui isocam filters}
The filter transmission for the {\it Spitzer}\ blue peak up, 16 \mic, filter (solid line)
compared to the ISOCAM LW3, 15 \mic, filter (dotted line).  For clarity, both filter curves
have been normalized to set their peak transmission to unity.
  }

\end{figure}

\begin{figure}[t*]

\plotone{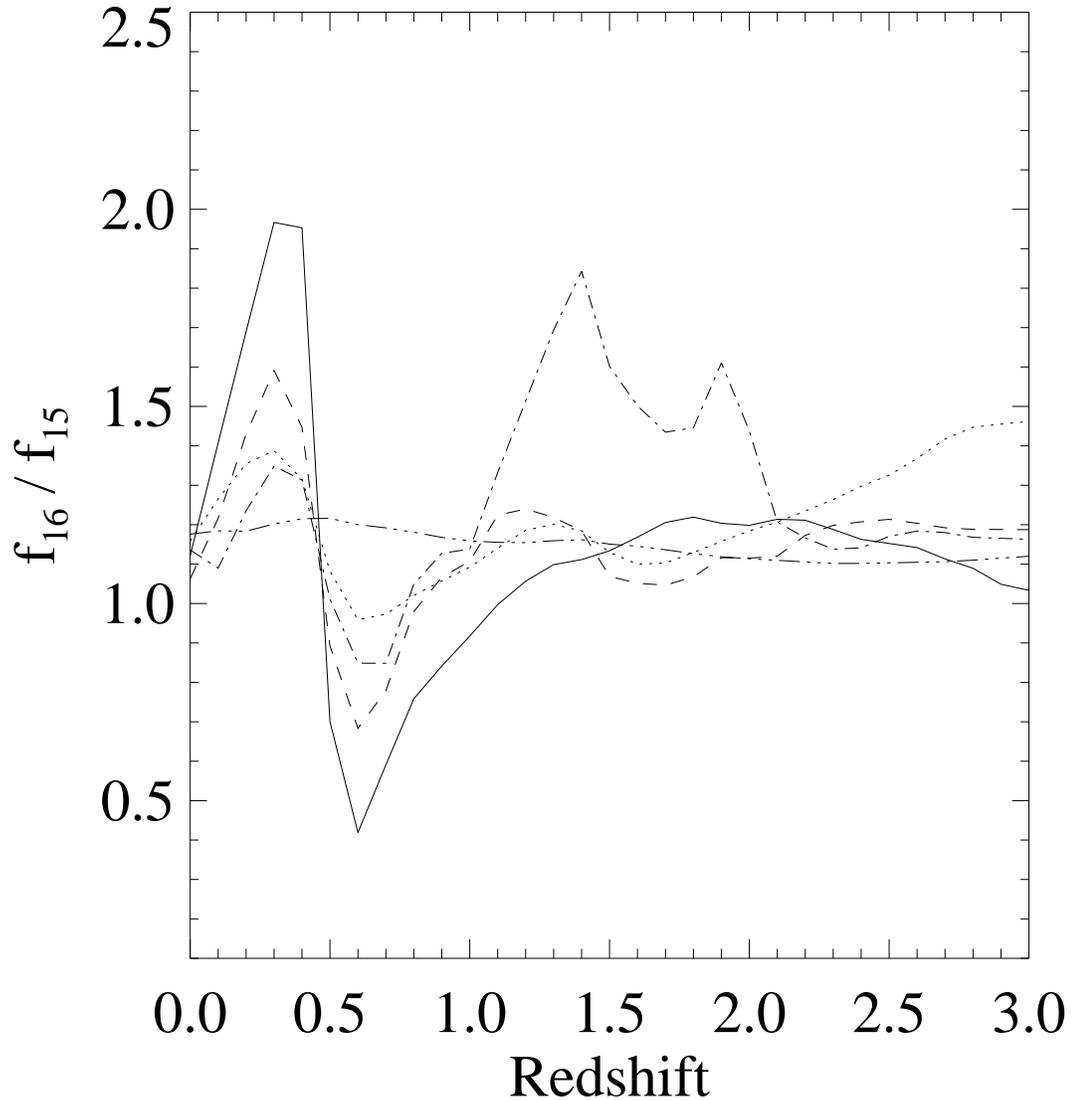}
\caption{\label{fig: pui isocam ratio}
  The predicted ratio of the IRS 16$\mu$m to ISOCAM 15 $\mu $m
  (LW3) flux densities as a function of redshift based on IRS spectra
  of template galaxies. We use the extreme silicate-absorption galaxy
  F00183-7111 \citep[solid line]{Spoon 2004}, UGC5101, a ULIRG with considerable
  9.7$\mu$m absorption \citep[dashed line]{Armus 2004}, the prototypical AGN Mrk231
  \citep[dotted line]{Weedman05}, the typical quasar PG1501+106 from
     \citep[tripple-dot-dashed line]{Hao05}
  and the average mid-IR SED of all starburst galaxies in the IRS GTO
  program from \citep[dot-dashed line]{Brandl05}. }

\end{figure}

We compare our 16 $\mu$m photometry directly with ISOCAM 15 $\mu$m
observations of the same field.  Figure \ref{fig: iso compare}\ shows
the object by object flux comparison.  We have applied two color
correction terms to the {\it Spitzer}\ fluxes.  First, there is the
small correction for the difference in the photometric system (see
above).  Secondly, there is a relatively large correction necessary
due to the difference in the effective wavelength of the filters.
Figure \ref{fig: pui isocam filters}\ shows the filter transmission
for the {\it Spitzer}\ and ISOCAM filters. If we define the effective
wavelength following \cite{Reach 2005}, we find $\lambda_{eff} =
15.5$\ for the PUI filter.  The ISOCAM LW3 photometry uses
$\lambda_{eff} = 14.3$\footnote{See the ``ISOCAM Photometry Report'',
1998;
http://www.iso.vilspa.esa.es/users/expl\_lib/CAM/photom\_rep\_fn.ps.gz} A
red source (e.g.\ $f_{\nu}\sim \lambda^2$) will have $\sim 15$\%\ more
flux at the redder wavelength, so we adopt that as our color
correction in the figure.  In general, sources agree within the
uncertainties, though the {\it Spitzer} fluxes may be systematically
high by $\sim 10-20$\%.  This disagreement may be attributable to the
difference in the filter bandpasses between IRS and ISOCAM, as
prominent features move into and out of the filters at different
redshifts.  For example, at redshift $z\sim 0.4$, the 9.7 \mic\ 
silicate absorption trough will affect the bluer LW3 filter more than
the PUI filter.  Figure \ref{fig: pui isocam ratio}\ shows the ratio of flux
density in the two filters as a function of redshift for a variety of
template sources.  As can be seen in the figure, a factor of two is
not unrealistic, so the difference in individual objects is
reasonable.  The four brightest objects in Figure \ref{fig:
  iso compare}\ are at redshifts 0.41, 0.45, 1.2, and 2.0.

\begin{figure}[t*]

\plotone{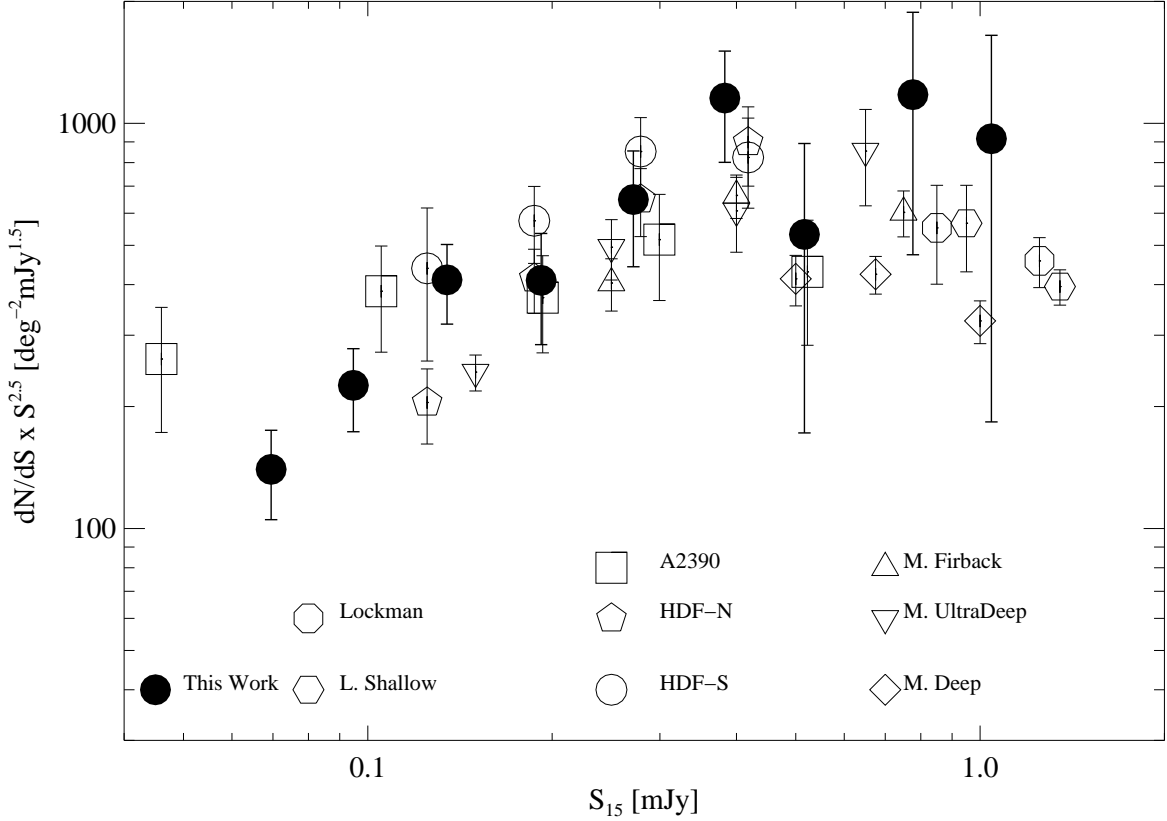}
\caption{\label{fig: nc}
  The differential 16 $\mu$m galaxy number counts measured by {\it Spitzer}\ 
  and ISO.  The Euclidean slope has been removed.  Poissonian error
  bars are shown.  ISOCAM points include: HDF-N, HDF-S, and the Marano
  surveys \citep[][ and the references therein]{Elbaz 1999}; the
  gravitational lensing cluster survey \citep{Altieri 1999}; ELAIS-S1
  \citep{Gruppioni 2002}; and the Lockman and Lockman Shallow surveys
  \citep{Rodighiero 2004}.  }

\end{figure}

We also compare the differential galaxy-number counts of our survey to
those measured by ISO (Figure \ref{fig: nc}).  We use the exposure map
to define the area coverage and depth.  For this initial comparison,
we do not apply an incompleteness correction.  The use of the
ultradeep IRAC data ensures that there is little contamination from
spurious objects. We see general agreement of the counts within the
large error bars.  It is important to highlight this agreement,
because of the use of ISOCAM number counts as a benchmark for
comparison to {\it Spitzer}\ surveys.  Substantial differences are seen
between the first MIPS 24$\mu$m data \citep{Marleau 2004, Papovich
  2004}\ and the ISOCAM data.  {\it Spitzer}\ 24 \mic\ counts appear to peak at fainter
fluxes than those measured by ISO at 15 \mic.  \cite{Gruppioni 2005}\ attribute
this difference to a population of starburst galaxies seen by {\it Spitzer}\ 
at higher redshifts than those which contribute to the ISO counts.  We
conclude that the differences reported between 24 $\mu$m and 15 $\mu$m
number counts are real and not the result of systematic
observatory-based bias.  However, we emphasize that our {\it Spitzer}\ counts
are derived from the same field as the faint end of the ISO counts, so
we do not remove any effects of cosmic variance.

\section{Discussion}
\label{sec: bolcor}

The mid-infrared detection of sources with known redshifts enables an
estimate to be made of their bolometric luminosity. This is possible
because of the empirical correlations seen between the mid- and
far-infrared luminosities of IRAS detected galaxies in the local
Universe \citep{Chary and Elbaz 2001}. To estimate the bolometric
correction of galaxies in our sample, we used a sample of 5 local
objects with mid-infrared and far-infrared SEDs accurately determined
from {\it Spitzer}\ IRS and IRAS. These are: M51 \citep[from the SINGS
legacy project first data release]{Kennicutt
  2003}\footnote{http://ssc.spitzer.caltech.edu/legacy/}; F00183-7111
\citep{Spoon 2004}, UGC5101, Mrk1014, Mrk463 \citep{Armus 2004}; and 
NGC7714 \citep{Brandl 2004}. We integrated the mid-infrared spectrum
and the IRAS broadband photometry \citep{Soifer 1989}\ with a dual temperature dust model
to obtain a complete SED from 5-1000$\mu$m.

At the redshift of each 16$\mu$m source, we convolved the redshifted
template SEDs through the IRS 16$\mu$m and MIPS 24$\mu$m bandpasses.
We selected the template whose 16/24 flux ratio was closest to that
observed.  The selected template was then scaled to match the observed
16 and 24 micron fluxes using weights which are the inverse of the
fractional flux uncertainty. The statistical uncertainties associated
with this process are directly proportional to the flux uncertainty
which is small due to the high S/N of the GOODS 24$\mu$m data. The
systematic uncertainties are substantial since they rely on the
assumption that high-redshift infrared SEDs are similar to that of
local galaxies and the fact that the galaxy SEDs chosen here do not
span the whole range of dust templates \citep[see e.g.][Marcillac et
al.\ 2005, in prep.]{Chary and Elbaz 2001}.

At $z>1.3$, the rest-frame 9.7 $\mu$m silicate absorption feature
enters the MIPS 24 $\mu$m wavelength range, and it is likely that
varying absorption strength would cause increased scatter in the
inferred infrared luminosity (\lir).  The current sample includes too few objects at that
redshift to confirm that expectation.

\begin{figure}[t*]

\plotone{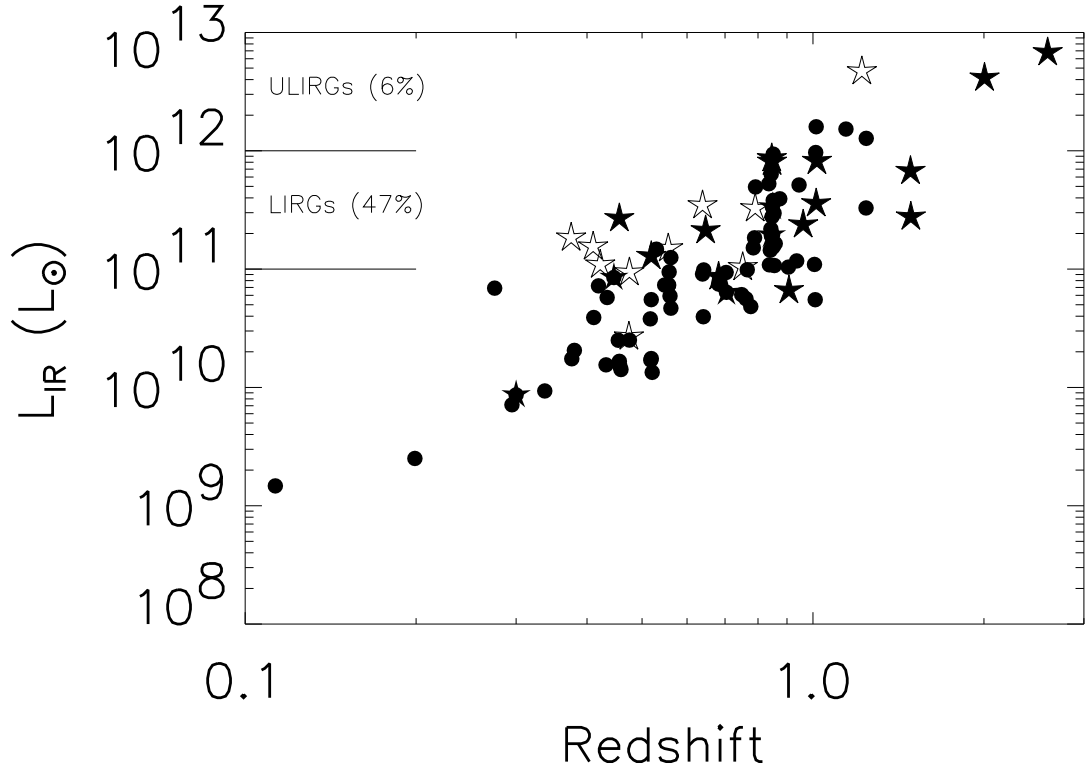}
\caption{\label{fig: LIR}
  The inferred \lir\ as a function of redshift for 16 \mic\ detected
  sources.  Open stars indicates soft-band-only {\it Chandra}\ 
  detections, and filled stars indicate hard-band detections.  All
  other sources are shown as filled circles.  The error in inferred
  \lir\ may be dominated by the template fitting, so propagated
  photometric errors are not shown.  }

\end{figure}

Figure \ref{fig: LIR}\ shows the inferred \lir\ for the sources in the
16 \mic\ catalog with spectroscopic redshifts.  More than half of the
16 $\mu$m sources (53 of 89) have inferred \lir\ greater than
$>10^{11}~L_{\odot}$, implying that they are in the LIRG class, and
six of them have ULIRG luminosities.  The detection of substantial
numbers of sub-LIRG luminosity objects at $z\le 1$, and sub-ULIRG
objects at $z<1.5$, in a few minutes of on-source integration
demonstrates the power of the 16 \mic\ observing mode.  We note that
the inference of \lir\ relies on spectroscopic redshifts, which by 
necessity are biased to the brighter optical sources.  Some extremely
red, high luminosity sources could be missed.  In the particular case
of {\it Chandra}\ sources, 75\% (25/33) have spectroscopic redshifts, 
so the bias introduced is likely to be small.

Most of the {\it Chandra}-detected sources are among the luminous objects,
including three of the six ULIRGs. Two of the ULIRGs are detected in
the hard band (2-8 keV), implying at least a contribution from an
active nucleus.  This result is consistent with existing trends in the
literature for ULIRG class objects \citep{Tran 2001}.  However, we
find substantial evidence for AGN activity (and potentially AGN
dominated luminosity) below $10^{12}~\mbox{L}_{\odot}$.

\begin{figure}[t*]

\plotone{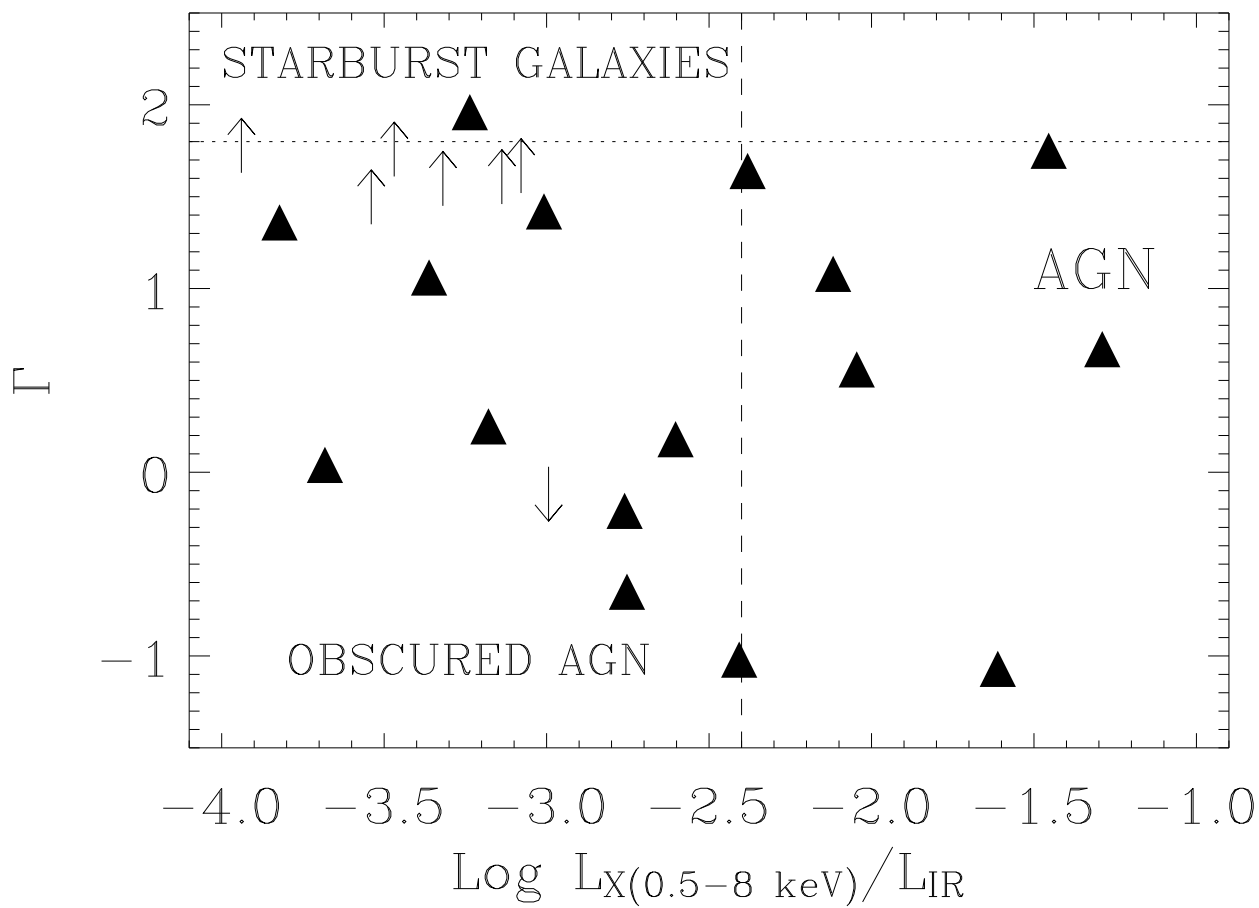}
\caption{\label{fig: xlir}
  The ratio of X-ray to infrared luminosities vs. the photon index.
  $L_{\rm IR}$\ were calculated as described in the text.  $L_X$\ were
  calculated using the photon index to extrapolate to the rest-frame
  full-band (0.5-8 keV).  Sources without hard-band detections are
  shown as lower limits.  The vertical dashed line shows the ratio of
  strong AGN limit of 0.004, and the horizontal line shows the local
  Seyfert $\Gamma$\ ratio of 1.8. }

\end{figure}

Because the X-ray data in the 2 Ms field is very deep, not all {\it Chandra}\ 
sources at these depths are necessarily AGN dominated sources.
\cite{Fadda 2002}\ conclude that only 5 of their 16 MIR-detected X-ray
sources are unambiguously AGN dominated.  In the current sample, we
identify 21 hard band (2-8 keV) detections.  These galaxies very
likely host an AGN, but without a more complete sampling of their SED
we cannot definitively determine the AGN contribution to the
bolometric luminosity.  We examine the ratio of X-ray to infrared
luminosities in Figure \ref{fig: xlir}.  \cite{Alexander 2005}\ find
that most AGN-dominated luminous IR galaxies have ${{L_X}\over{L_{\rm
      FIR}}} > 0.004$, but that some AGN sources have ratios as much
as ten times lower.  They correct the $L_X$\ for extinction, while we
do not; increasing the X-ray flux will only lead to more AGN
dominated sources, so our measurement is a lower limit.  Alexander et
al.\ also find that sub-millimeter selected sources classified as AGN
have photon indices $\Gamma \simeq 1.8$, consistent with local
Seyferts.  Lower $\Gamma$\ values would imply stronger AGN
contribution to the X-ray flux.  The 16 \mic\ sources with
hard-band {\it Chandra}\ counterparts generally meet one or both of these
criteria.  The objects most likely to be starbursts (low X-ray to IR
ratio and high values of $\Gamma$) are mostly soft-band only sources.
The same conclusion is reached using the ratio of infrared to X-ray
fluxes, $IR/X$\ (Weedman et al.\ 2004).

The 16 \mic\ detected X-ray sources have approximately the same
distribution of X-ray flux as the non-detections.  A marginal
difference is seen in their photon indices, with the MIR sample having
a median $\Gamma$\ value of 1.1 compared to 1.4 for the
non-detections.  This difference could indicate a higher fraction of
AGN in the MIR sample.  The same fraction ($\sim 50$\%) of each sample
has no $\Gamma$\ value due either to the faintness of the full-band
X-ray flux or to a non-detection in one of the X-ray bands.

\subsection{Evidence for PAH emission}

The ratio of 16 to 24 \mic\ flux density is expected to be a strong
function of redshift for objects with substantial features in their
MIR spectra.  These spectra are a combination of: a (usually red)
continuum slope; emission from PAH molecules at 6.2, 7.7, 8.6, 11.3,
and 12.7 \mic; and a potentially deep silicate absorption trough at
9.7 \mic.  \cite{Takagi and Pearson 2005}\ demonstrate the utility of
the 16/22 \mic\ ratio as a crude redshift indicator (the 22 and 24
\mic\ filters are similar).  The relatively broad width of the {\it
  Spitzer}\ bandpasses complicates the interpretation of the flux
ratio, because some redshifts will place both absorption and emission
features within the filter.

\begin{figure}[b*]

\plotone{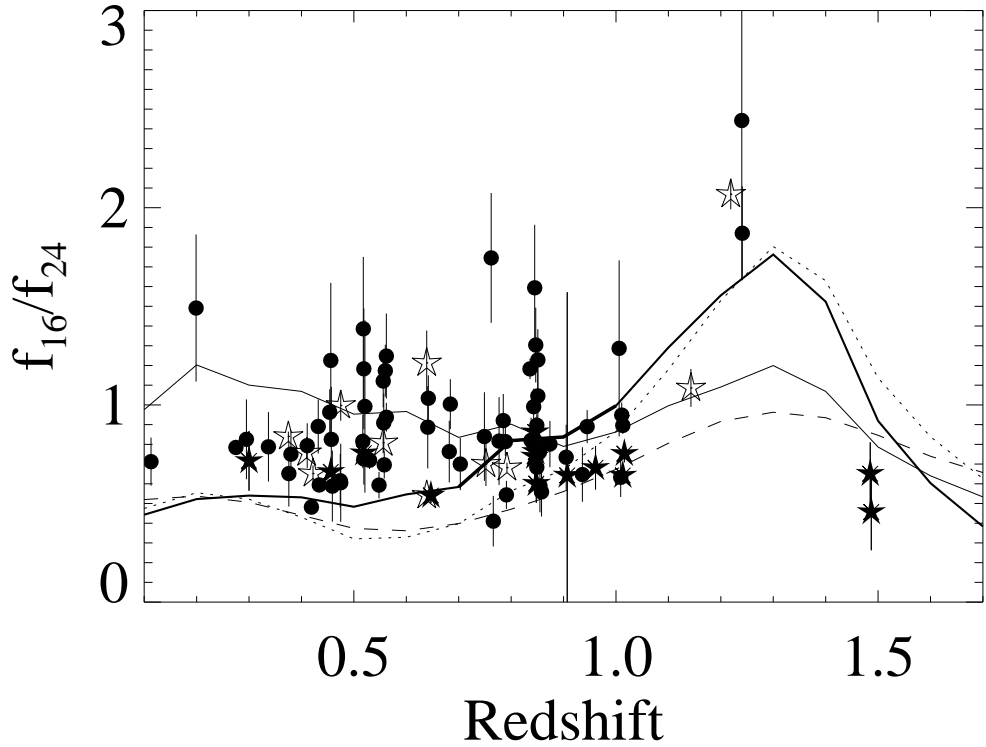}
\caption{\label{fig: flux ratio}
  The ratio of 16 to 24 $\mu$m flux density, $f_{\nu}$, as a function
  of redshift.  Star symbols indicate sources with {\it Chandra}\ 
  counterparts, filled stars for hard band detections.  Template
  ratios are shown for a PAH-dominated SED \citep[M51; solid line; SINGS DR1;][]{Kennicutt 2003}
, an edge-on starburst \citep[M82; thick line;][]{Forster
    Schreiber 2001}, a ULIRG \citep[U5101; dotted line;][]{Armus 2004},
  and an AGN \citep[Mrk 231; dashed line;][]{Weedman05}.  }

\end{figure}

In Figure \ref{fig: flux ratio}, we plot the ratio of 16 to 24 $\mu$m
flux density, $f_{\nu}$, for the current sample and the expected ratio
for local templates, for comparison.  We find most sources have flux
density ratios of $\sim 0.75$, but that higher ratios are observed at
some redshifts.  These redshifts are the ones where we expect the
redshifting of PAH features to increase the 16 $\mu$m flux density
relative to the MIPS 24 $\mu$m flux density.  In particular, at
redshifts near $z\sim 0.5$, the 11.3 $\mu$m PAH feature is in the
middle of IRS 16 $\mu$m band.  This feature can have low equivalent
width in objects with a strong VSG continuum, such as some ULIRGs, but
may dominate in relatively unextinguished starbursts like M51.  At
redshift $z\sim 0.8$, the 7.7 $\mu$m PAH complex starts to shift into
the IRS 16 $\mu$m bandpass, as seen in the rising ratio for M82.  At
$z\sim 1.3$, two effects contribute to the 16/24 ratio: the silicate
absorption feature shifts into the MIPS 24 $\mu$m band, and the 6.2
and 7.7 \mic\ PAH features are centered in the blue PU bandpass.

It appears that most of the objects in the 16 \mic\ sample are
significantly bluer (that is have higher 16/24 ratios) than the ULIRG
template.  Despite having high luminosity PAH emission, the ULIRG
spectra also have a strong continuum so the PAH's have low equivalent
width (EW) and do not dominate the SED.  M51, on the other hand, is
dominated by PAH and [NeII] emission.  A few sources have even higher
16/24 ratios than M51, which could be explained by a bluer continuum
or PAH features with several times higher EWs.

Similar results were observed by \cite{Elbaz 2005}\ and \cite{Marcillac 
2005}\ in comparing ISOCAM 15 \mic\ fluxes to MIPS 24
\mic\ photometry.  They find that the ratio of flux, $\nu f_{\nu}$,
for the two filters is consistent with the expected SED for
star-forming objects with significant PAH emission.  Marcillac et al.\ 
examine the ISO HDF-N data, with some sources overlapping the present
sample.  Their independent measurement also suggests evidence for PAH
emission at $z\sim 1$.

We also note that the presence of PAH features helps to distinguish
AGN from starburst dominated sources, in sources with weak silicate
absorption.  \cite{Charmandaris 2004}\ suggest that, in fact, the 16
to 24 $\mu$m flux density ratio can differentiate between the two SED
types in the case of SCUBA-selected sources.  We find that the present
survey shows a low 16/24 ratio for sources with {\it Chandra}\ hard band
counterparts.  These sources have the ratio expected from the Mrk 231
template.  Nonetheless, highly extincted, X-ray weak AGN may still 
contaminate the starburst portion of the sample.  To fully explore this
issue, substantial MIR spectroscopic data may be required.

\section{Summary and Conclusions}

We have presented a pilot study of 16 $\mu$m imaging of faint
extragalactic objects with the {\it Spitzer}\ IRS.  Our photometric
results show good agreement with the 15 $\mu$m ISO survey of the same
area.  We find evidence of PAH emission at $z\sim 0.5$\ and $z>0.8$\ 
in the ratio of 16 to 24 $\mu$m fluxes.  The scatter in the flux ratio
is large at these redshifts, potentially indicating a broad range of
sources with differing emission properties.  In principle, we may be
able to use photometric surveys such as this one to constrain the
strength of the PAH emission relative to the continuum.  If so, it may
be possible to calibrate the 16 \mic\ flux as a measurement of star
formation at $z\sim 1$.  However, such inferences will depend on a
good understanding of how the strength of the PAH features varies, as
well as a good baseline for the underlying continuum.

These results suggest that larger surveys with the 16 $\mu$m filter
will detect PAH emission at $z\sim 1$\ in statistically significant
samples.  The flux limits achievable in a few minutes of observation
with the PUI mode will detect sources much fainter than the
spectroscopic limit.  These observations will enable direct comparison
of photometrically measured PAH emission at $z\sim 1$\ to those
detected at $z\sim 2$\ with MIPS.

\acknowledgements

We thank M. E. Dickinson for access to the pre-release GOODS MIPS
photometry.   This work is based in part on observations
made with the {\it Spitzer Space Telescope}, which is operated by the Jet
Propulsion Laboratory, California Institute of Technology under NASA
contract 1407. Support for this work was provided by NASA through an
award issued by JPL/Caltech.

\clearpage

\begin{deluxetable}{rrrrrrrrr}
\tabletypesize{\scriptsize}
\tablecaption{Photometry\label{tab: catalog}}
\tablehead{
\colhead{R.A. (J2000)} &
\colhead{Dec. (J2000)} &
\colhead{$f_{16}$\ ($\mu$Jy)} &
\colhead{$\sigma_{16}$\ ($\mu$Jy)} &
\colhead{$f_{24}$\ ($\mu$Jy)} &
\colhead{$\sigma_{24}$\ ($\mu$Jy)} &
\colhead{$z$} &
\colhead{$z$\ ref\tablenotemark{a}} &
\colhead{X-ray ID\tablenotemark{b}} 
}

\startdata

           189.29108  &             62.13359  &   139  &    24  &   159  &     8  &    \nodata  &    \nodata  &        360  \\ 
           189.23334  &             62.13561  &   568  &    20  &   832  &     7  &      0.792  &          1  &        288  \\ 
           189.24095  &             62.14082  &   176  &    19  &   150  &     5  &      0.560  &          2  &    \nodata  \\ 
           189.22546  &             62.14306  &   117  &    18  &   258  &     5  &    \nodata  &    \nodata  &        281  \\ 
           189.29332  &             62.14336  &   236  &    28  &   281  &     8  &    \nodata  &    \nodata  &    \nodata  \\ 
           189.29060  &             62.14479  &   878  &    32  &  1120  &    14  &    \nodata  &    \nodata  &        359  \\ 
           189.25629  &             62.14495  &   130  &    24  &   185  &     7  &      0.703  &          1  &    \nodata  \\ 
           189.20917  &             62.14573  &   348  &    20  &   585  &     8  &      0.434  &          1  &    \nodata  \\ 
           189.27412  &             62.14622  &   110  &    16  &    83  &     6  &    \nodata  &    \nodata  &    \nodata  \\ 
           189.22844  &             62.14643  &   134  &    17  &   246  &     6  &      0.790  &          1  &    \nodata  \\ 
           189.30330  &             62.14756  &   205  &    24  &   273  &     7  &    \nodata  &    \nodata  &    \nodata  \\ 
           189.18565  &             62.14861  &    79  &    19  &    98  &     5  &    \nodata  &    \nodata  &    \nodata  \\ 
           189.29300  &             62.14959  &   395  &    26  &   407  &     9  &    \nodata  &    \nodata  &    \nodata  \\ 
           189.19591  &             62.15186  &    66  &    14  &    90  &     6  &      0.905  &          1  &    \nodata  \\ 
           189.17033  &             62.15443  &    99  &    17  &    71  &     5  &    \nodata  &    \nodata  &    \nodata  \\ 
           189.23238  &             62.15482  &   408  &    21  &   846  &    10  &      0.419  &          1  &    \nodata  \\ 
           189.28125  &             62.15555  &   128  &    24  &   128  &     8  &    \nodata  &    \nodata  &    \nodata  \\ 
           189.15843  &             62.15610  &    87  &    27  &   212  &     6  &      0.766  &          1  &    \nodata  \\ 
           189.18608  &             62.15747  &    65  &    16  &    47  &     5  &    \nodata  &    \nodata  &    \nodata  \\ 
           189.19670  &             62.15831  &    65  &    14  &    53  &     8  &    \nodata  &    \nodata  &    \nodata  \\ 
           189.20271  &             62.15891  &   104  &    16  &   127  &     7  &      0.517  &          1  &    \nodata  \\ 
           189.17740  &             62.15941  &   343  &    21  &   476  &     6  &      0.530  &          1  &    \nodata  \\ 
           189.16095  &             62.16030  &   172  &    25  &   214  &     6  &      0.873  &          1  &    \nodata  \\ 
           189.17416  &             62.16194  &    86  &    22  &   117  &     6  &      0.845  &          1  &        219  \\ 
           189.17314  &             62.16339  &   327  &    23  &   433  &     7  &      0.518  &          1  &        217  \\ 
           189.16887  &             62.16757  &    89  &    23  &   106  &     6  &      0.749  &          1  &    \nodata  \\ 
           189.14030  &             62.16826  &   438  &    25  &   581  &     9  &      1.016  &          1  &        177  \\ 
           189.15561  &             62.17088  &    78  &    21  &    46  &     5  &    \nodata  &    \nodata  &    \nodata  \\ 
           189.38908  &             62.17114  &   197  &    31  &   284  &    10  &    \nodata  &    \nodata  &    \nodata  \\ 
           189.12247  &             62.17173  &   158  &    18  &   167  &     5  &    \nodata  &    \nodata  &    \nodata  \\ 
           189.30412  &             62.17455  &   102  &    22  &   183  &     6  &      0.858  &          1  &    \nodata  \\ 
           189.28342  &             62.17473  &   310  &    22  &   163  &     8  &    \nodata  &    \nodata  &    \nodata  \\ 
           189.21304  &             62.17522  &   745  &    29  &   984  &     9  &      0.410  &          1  &        265  \\ 
           189.36525  &             62.17670  &   878  &    37  &   742  &     8  &    \nodata  &    \nodata  &    \nodata  \\ 
           189.37094  &             62.17753  &   226  &    33  &   245  &     6  &      0.784  &          1  &    \nodata  \\ 
           189.13484  &             62.17790  &    99  &    18  &   118  &     6  &    \nodata  &    \nodata  &    \nodata  \\ 
           189.34918  &             62.17939  &   141  &    24  &   198  &     7  &      0.113  &          1  &    \nodata  \\ 
           189.12134  &             62.17940  &   465  &    25  &   724  &    12  &      1.013  &          1  &        158  \\ 
           189.19768  &             62.17942  &   300  &    29  &    65  &     7  &    \nodata  &    \nodata  &    \nodata  \\ 
           189.19420  &             62.18023  &   315  &    29  &   354  &     7  &      0.945  &          1  &    \nodata  \\ 
           189.13847  &             62.18055  &    66  &    22  &    51  &     5  &      1.006  &          1  &    \nodata  \\ 
           189.28607  &             62.18078  &   193  &    27  &   243  &     6  &      0.411  &          1  &    \nodata  \\ 
           189.36378  &             62.18087  &   127  &    27  &    62  &     6  &    \nodata  &    \nodata  &    \nodata  \\ 
           189.21376  &             62.18095  &   155  &    30  &    70  &     6  &    \nodata  &    \nodata  &        266  \\ 
           189.30556  &             62.18172  &    74  &    15  &   114  &     6  &      0.936  &          1  &    \nodata  \\ 
           189.14169  &             62.18184  &   115  &    20  &    66  &     5  &      0.762  &          1  &    \nodata  \\ 
           189.38663  &             62.18212  &    90  &    21  &   147  &     8  &    \nodata  &    \nodata  &    \nodata  \\ 
           189.28464  &             62.18222  &   423  &    26  &   648  &     7  &      0.422  &          1  &        353  \\ 
           189.39809  &             62.18230  &   488  &    21  &   354  &     7  &    \nodata  &    \nodata  &    \nodata  \\ 
           189.12839  &             62.18300  &    88  &    21  &   106  &     6  &      0.295  &          1  &    \nodata  \\ 
           189.01353  &             62.18633  &   655  &    33  &  1210  &     9  &      0.639  &          1  &         67  \\ 
           189.13100  &             62.18708  &   430  &    20  &   480  &     6  &      1.013  &          1  &    \nodata  \\ 
           189.38336  &             62.18852  &   260  &    29  &   235  &     6  &    \nodata  &    \nodata  &    \nodata  \\ 
           189.17239  &             62.19147  &   232  &    17  &   287  &     7  &    \nodata  &    \nodata  &    \nodata  \\ 
           189.32979  &             62.19196  &   213  &    24  &   190  &     6  &      0.556  &          1  &    \nodata  \\ 
           189.16946  &             62.19321  &    68  &    12  &   110  &     5  &    \nodata  &    \nodata  &    \nodata  \\ 
           189.31349  &             62.19333  &    96  &    21  &    99  &     5  &    \nodata  &    \nodata  &    \nodata  \\ 
           189.05197  &             62.19450  &   973  &    27  &  1240  &    13  &      0.275  &          1  &    \nodata  \\ 
           189.01643  &             62.19463  &   166  &    26  &    69  &     5  &    \nodata  &    \nodata  &    \nodata  \\ 
           189.02048  &             62.19480  &   137  &    25  &    59  &     4  &    \nodata  &    \nodata  &    \nodata  \\ 
           189.19241  &             62.19494  &   275  &    18  &   290  &     5  &      1.011  &          1  &    \nodata  \\ 
           189.17258  &             62.19509  &   134  &    15  &   225  &     5  &      0.548  &          1  &    \nodata  \\ 
           189.29378  &             62.19510  &   104  &    23  &   177  &     5  &      0.855  &          1  &    \nodata  \\ 
           189.03665  &             62.19544  &   418  &    27  &   265  &     6  &    \nodata  &    \nodata  &    \nodata  \\ 
           189.17992  &             62.19660  &    72  &    11  &   113  &     4  &      1.009  &          1  &    \nodata  \\ 
           189.02428  &             62.19663  &   109  &    26  &   167  &     4  &      1.484  &          2  &         74  \\ 
           189.15739  &             62.19703  &   142  &    20  &   173  &     4  &      0.838  &          1  &    \nodata  \\ 
           189.15952  &             62.19740  &   189  &    19  &   230  &     3  &      0.841  &          1  &    \nodata  \\ 
           188.93968  &             62.19881  &   150  &    21  &   225  &     8  &    \nodata  &    \nodata  &    \nodata  \\ 
           189.15422  &             62.19989  &    65  &    17  &    79  &     6  &      0.777  &          1  &    \nodata  \\ 
           189.02254  &             62.20059  &    87  &    27  &    71  &     6  &      0.456  &          1  &    \nodata  \\ 
           189.17482  &             62.20144  &    74  &    11  &    83  &     5  &      0.432  &          1  &    \nodata  \\ 
           189.14368  &             62.20358  &   853  &    32  &  1290  &     9  &      0.456  &          1  &        180  \\ 
           189.15337  &             62.20359  &   300  &    22  &   379  &     5  &      0.844  &          1  &    \nodata  \\ 
           189.16554  &             62.20385  &    26  &     9  &    28  &     5  &    \nodata  &    \nodata  &    \nodata  \\ 
           189.17880  &             62.20448  &   110  &    13  &   114  &     5  &      0.454  &          1  &    \nodata  \\ 
           188.98366  &             62.20535  &   295  &    25  &   430  &     6  &    \nodata  &    \nodata  &         48  \\ 
           189.21553  &             62.20573  &    90  &    19  &   126  &     6  &      0.300  &          1  &        267  \\ 
           189.14581  &             62.20672  &   315  &    24  &   336  &     6  &      0.562  &          1  &    \nodata  \\ 
           189.16486  &             62.20842  &    30  &     9  &    44  &     5  &    \nodata  &    \nodata  &        207  \\ 
           189.15422  &             62.20854  &   146  &    16  &   232  &     4  &    \nodata  &    \nodata  &    \nodata  \\ 
           189.20926  &             62.21098  &    73  &    16  &   119  &     7  &      0.474  &          1  &    \nodata  \\ 
           189.14377  &             62.21138  &   923  &    32  &   446  &     5  &      1.219  &          1  &        182  \\ 
           189.15680  &             62.21138  &    58  &    18  &    70  &     4  &      0.457  &          1  &    \nodata  \\ 
           189.13190  &             62.21203  &    95  &    21  &   109  &     5  &    \nodata  &    \nodata  &    \nodata  \\ 
           189.13654  &             62.21220  &   141  &    23  &   113  &     6  &      0.562  &          1  &    \nodata  \\ 
           189.16635  &             62.21386  &   425  &    20  &   493  &     6  &      0.846  &          1  &        211  \\ 
           189.18327  &             62.21389  &   343  &    22  &   424  &     5  &      0.556  &          1  &        227  \\ 
           189.17671  &             62.21447  &    64  &    17  &    26  &     5  &      1.240  &          2  &    \nodata  \\ 
           189.22455  &             62.21495  &   207  &    22  &   200  &     6  &      0.642  &          1  &    \nodata  \\ 
           189.16197  &             62.21589  &   265  &    22  &   244  &     7  &      1.143  &          1  &        203  \\ 
           189.20685  &             62.21594  &    66  &    21  &   109  &     6  &      0.475  &          1  &    \nodata  \\ 
           189.14230  &             62.21824  &   176  &    24  &   135  &     6  &    \nodata  &    \nodata  &    \nodata  \\ 
           189.14310  &             62.22007  &   137  &    26  &    86  &     5  &      0.845  &          1  &    \nodata  \\ 
           189.20706  &             62.22025  &   370  &    27  &   371  &    10  &      0.475  &          1  &        260  \\ 
           189.21022  &             62.22108  &   241  &    30  &   196  &     6  &      0.851  &          1  &    \nodata  \\ 
           189.14000  &             62.22213  &   320  &    29  &   323  &     7  &      0.843  &          1  &    \nodata  \\ 
           189.21275  &             62.22227  &   118  &    28  &    79  &     6  &      0.199  &          1  &    \nodata  \\ 
           189.20445  &             62.22270  &   104  &    32  &    48  &     8  &    \nodata  &    \nodata  &    \nodata  \\ 
           189.21581  &             62.23160  &   185  &    30  &   203  &     6  &      0.557  &          1  &    \nodata  \\ 
           189.19305  &             62.23459  &   144  &    23  &   211  &     5  &      0.961  &          1  &        240  \\ 
           189.20628  &             62.23516  &   130  &    20  &   186  &     6  &      0.752  &          1  &        258  \\ 
           189.17159  &             62.23911  &    60  &    14  &    51  &     6  &      0.519  &          1  &    \nodata  \\ 
           189.16225  &             62.23990  &    82  &    17  &    83  &     6  &    \nodata  &    \nodata  &    \nodata  \\ 
           189.14830  &             62.23993  &   615  &    32  &  1480  &    10  &      2.005  &          1  &        190  \\ 
           189.20126  &             62.24065  &   283  &    20  &   460  &     6  &    \nodata  &    \nodata  &        251  \\ 
           189.12537  &             62.24105  &    94  &    22  &   183  &     5  &    \nodata  &    \nodata  &    \nodata  \\ 
           189.14070  &             62.24189  &   137  &    23  &   188  &    10  &      0.519  &          1  &    \nodata  \\ 
           189.14941  &             62.24327  &   179  &    25  &   195  &    15  &    \nodata  &    \nodata  &    \nodata  \\ 
           189.13786  &             62.24362  &   107  &    19  &   109  &     6  &    \nodata  &    \nodata  &    \nodata  \\ 
           189.12784  &             62.24423  &    71  &    16  &    55  &     6  &    \nodata  &    \nodata  &    \nodata  \\ 
           189.19092  &             62.24640  &    79  &    19  &   165  &     7  &    \nodata  &    \nodata  &    \nodata  \\ 
           189.19527  &             62.24641  &   193  &    17  &   277  &    11  &      0.558  &          1  &    \nodata  \\ 
           189.18340  &             62.24736  &    64  &    15  &   138  &     6  &    \nodata  &    \nodata  &    \nodata  \\ 
           189.18481  &             62.24797  &    37  &    16  &    82  &     6  &      1.487  &          1  &        229  \\ 
           189.16055  &             62.24860  &    75  &    15  &    82  &     6  &    \nodata  &    \nodata  &    \nodata  \\ 
           189.17474  &             62.24905  &    80  &    13  &   117  &     6  &      0.849  &          2  &    \nodata  \\ 
           189.17014  &             62.25126  &    65  &    15  &    99  &     6  &    \nodata  &    \nodata  &    \nodata  \\ 
           189.10291  &             62.25286  &    75  &    17  &    84  &     6  &      0.641  &          1  &    \nodata  \\ 
           189.13535  &             62.25359  &   143  &    17  &   142  &     7  &      0.684  &          1  &    \nodata  \\ 
           189.13789  &             62.25376  &    41  &    17  &    42  &     7  &      0.521  &          1  &    \nodata  \\ 
           189.10132  &             62.25700  &   116  &    23  &   152  &     6  &      0.682  &          1  &    \nodata  \\ 
           189.16545  &             62.25726  &   121  &    19  &   161  &     6  &      0.380  &          1  &    \nodata  \\ 
           189.09558  &             62.25735  &   335  &    29  &   529  &     6  &      2.590  &          1  &        137  \\ 
           189.19266  &             62.25759  &   433  &    27  &   544  &     8  &      0.851  &          1  &    \nodata  \\ 
           189.07227  &             62.25818  &   448  &    27  &   499  &     7  &      0.849  &          1  &    \nodata  \\ 
           188.98537  &             62.25836  &   248  &    24  &   237  &    10  &    \nodata  &    \nodata  &    \nodata  \\ 
           189.01689  &             62.25971  &    90  &    20  &    87  &     6  &    \nodata  &    \nodata  &    \nodata  \\ 
           188.99142  &             62.26019  &   653  &    35  &   925  &    11  &    \nodata  &    \nodata  &    \nodata  \\ 
           189.16722  &             62.26089  &    94  &    20  &    81  &     7  &    \nodata  &    \nodata  &    \nodata  \\ 
           189.03229  &             62.26176  &   137  &    24  &   101  &     6  &    \nodata  &    \nodata  &    \nodata  \\ 
           189.09373  &             62.26234  &   390  &    28  &   721  &     7  &      0.647  &          1  &        132  \\ 
           189.08890  &             62.26270  &   210  &    25  &   112  &     5  &      1.241  &          2  &    \nodata  \\ 
           189.14543  &             62.26355  &    79  &    17  &   100  &     5  &      0.337  &          1  &    \nodata  \\ 
           189.00925  &             62.26373  &   214  &    28  &   176  &     7  &    \nodata  &    \nodata  &         62  \\ 
           188.99887  &             62.26376  &  1235  &    34  &  1470  &    13  &      0.375  &          1  &         57  \\ 
           189.07265  &             62.26416  &    95  &    23  &   145  &    10  &      0.375  &          2  &    \nodata  \\ 
           189.01863  &             62.26440  &   216  &    20  &   102  &     6  &    \nodata  &    \nodata  &    \nodata  \\ 
           189.03365  &             62.26478  &    74  &    22  &   125  &     7  &      0.459  &          1  &    \nodata  \\ 
           189.07832  &             62.26699  &   176  &    25  &   150  &     5  &    \nodata  &    \nodata  &    \nodata  \\ 
           189.13182  &             62.26773  &   245  &    22  &   301  &     5  &      0.788  &          1  &    \nodata  \\ 
           189.15320  &             62.26792  &    85  &    17  &    81  &     5  &      0.851  &          1  &    \nodata  \\ 
           189.12204  &             62.27040  &   125  &    20  &   208  &     5  &      0.847  &          1  &        160  \\ 
           189.01381  &             62.27181  &   126  &    18  &   262  &     9  &    \nodata  &    \nodata  &    \nodata  \\ 
           189.13525  &             62.27438  &   132  &    18  &   173  &     5  &      0.854  &          2  &    \nodata  \\ 
           189.14528  &             62.27447  &   368  &    21  &   482  &     5  &      0.847  &          1  &        187  \\ 
           189.16695  &             62.27649  &    86  &    22  &    65  &     5  &    \nodata  &    \nodata  &    \nodata  \\ 
           189.14203  &             62.27809  &    39  &    17  &    96  &     5  &    \nodata  &    \nodata  &    \nodata  \\ 
           189.14694  &             62.28175  &   152  &    17  &   173  &     5  &    \nodata  &    \nodata  &    \nodata  \\

\enddata

\tablenotetext{a}{1:  \cite{Cowie 2004}\ and the references therein;  2:  \cite{Wirth 2004}}  
\tablenotetext{b}{Catalog index from \cite{Alexander 2003}}

\end{deluxetable}

\clearpage

\begin{figure}[t*]

\plottwo{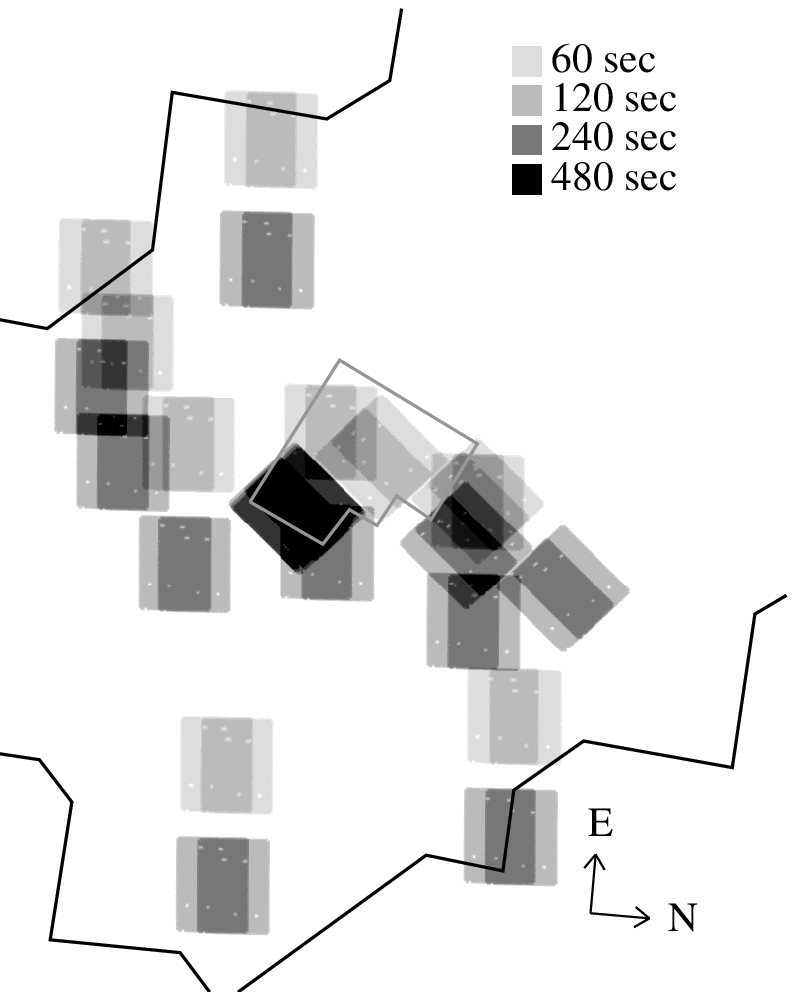}{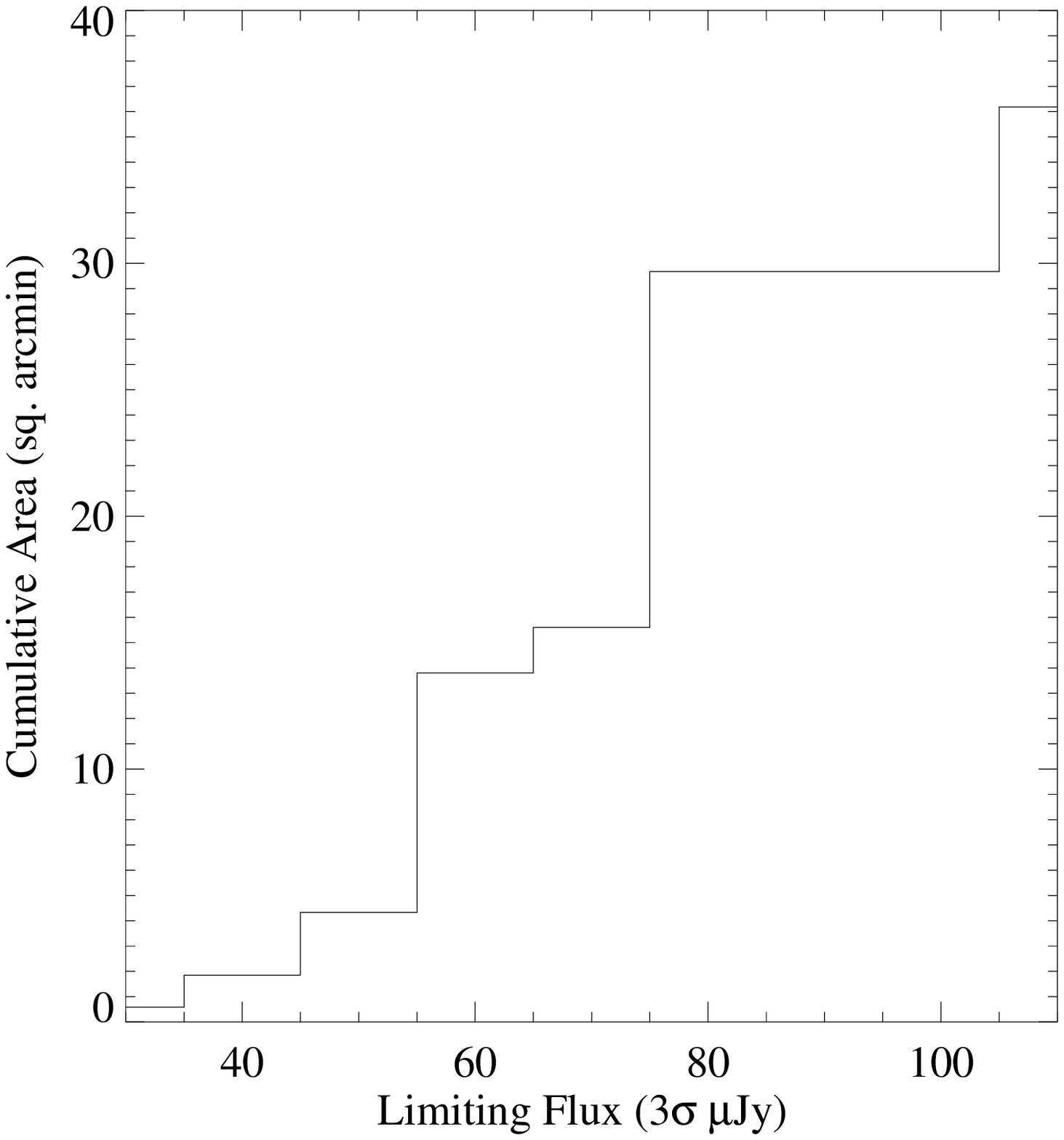}
\caption{ \label{fig: coverage}
(a) The coverage map for the 16 $\mu$m survey.  Integration times and
position angle are 
indicated.  The central chevron indicates the location of the HDF-N,
and the other solid lines denote the limits of the ACS GOODS survey.
(b) The cumulative area covered as a function of limiting flux.  We 
show that the majority of the area (30 of 35 square arcminutes) reach
a $3\sigma$\ sensitiviity of 75 $\mu$Jy or better.
  }

\end{figure}

\clearpage

\clearpage

\clearpage

\clearpage

\clearpage

\clearpage

\clearpage

\end{document}